\documentclass[reprint, prl]{revtex4-1}
	\usepackage{graphicx}
	\usepackage{amsmath}
	\usepackage{amsfonts}
	\usepackage{amssymb}
	\usepackage{color}
	\usepackage[colorlinks=true, citecolor=blue, urlcolor=blue, linkcolor=black]{hyperref}

\begin{document}
		\title{Detecting the ``phonon wind'' in superfluid \(\mathbf{^4He}\) by a nanomechanical resonator}
	
		\author{A.\,M. Gu\'{e}nault}
		\author{A. Guthrie}\email{a.guthrie1@lancaster.ac.uk}
		\author{R.\,P. Haley}
		\author{S. Kafanov}\email{sergey.kafanov@lancaster.ac.uk}
		\author{Yu.\,A. Pashkin}
		\author{G.\,R. Pickett}
		\author{V. Tsepelin}
		\author{D.\,E. Zmeev}
		
		\affiliation{Department of Physics, Lancaster University, Lancaster, LA1 4YB, United Kingdom}
	
		\author{E. Collin}
		\author{R. Gazizulin}
		\author{O. Maillet}
		\affiliation{Universit\'{e} Grenoble Alpes, CNRS Institut N\'{e}el, BP 166, 38042, Grenoble Cedex 9, France}
		
	     \begin{abstract}
	    	 Nanoscale mechanical resonators are widely utilized to provide high sensitivity force detectors. Here we demonstrate that such high quality factor resonators immersed in superfluid \(^4\mathrm{He}\) can be excited by a modulated flux of phonons. A nanosized heater immersed in superfluid \(^4\mathrm{He}\) acts as a source of ballistic phonons in the liquid -- ``phonon wind''. When the modulation frequency of the phonon flux matches the resonance frequency of the mechanical resonator, the motion of the latter can be excited. This ballistic thermomechanical effect can potentially open up new types of experiments in quantum fluids.
	     \end{abstract}
		
		\maketitle
	
		Nanoelectromechanical systems (NEMS) are finding an increasing range of application owing to their small size, high operating frequencies, high intrinsic quality factors \cite{ekinci2005nanoelectromechanical} and exceptionally small masses \cite{chaste2012nano} leading to very high force sensitivities \cite{melcher2014calibrating}. These properties make NEMS devices perfect candidates for studying quantum liquids, since a resonator with dimension \(50\,\mathrm{nm}\) can already probe the liquid properties on length scales comparable to the coherence lengths in superfluid \(^3\mathrm{He}\) \cite{vollhardt2013superfluid} and the de Broglie wavelengths of thermal excitations in superfluid \(^4\mathrm{He}\) at sub-mK temperatures \cite{guenault2003basic}. Liquid \(^4\mathrm{He}\) is the most investigated quantum fluid, with a well-understood spectrum of thermal excitations \cite{hohenberg2000microscopic} and topological defects \cite{tsubota2009quantum}. This makes it an ideal starting point for investigating the behavior of high-frequency nanomechanical resonators in superfluids. Moreover, recent theoretical and experimental research in optomechanics combined with superfluid \(^4\mathrm{He}\) \cite{lorenzo2014superfluid, romero2018quantum, macdonald2016optomechanics, lee2010cooling, Harris2016laser} makes nanomechanical resonators favorable candidates for such investigations.
	
		In this paper, we study the interaction between two doubly clamped nanomechanical beams mediated by thermal excitations in superfluid \(^4\mathrm{He}\). At the lowest temperatures, the only thermal excitations available in superfluid \(^4\mathrm{He}\) are phonons \cite{enss2005low}. We drive a nanobeam resonator by ``illuminating" it with a modulated flux of phonons generated by a heater in superfluid \(^4\mathrm{He}\). The phonon flux sensed by the nanobeam is analogous to the photon flux studied in the landmark experiment by P.\,N.\,Lebedev \cite{lebedew1901untersuchungen}, where the pressure of photons was measured.
	
		We operate two doubly clamped nanobeams placed side by side; one used off-resonance as a heater and the other as a detector. The nanobeams are made of silicon nitride (\(\mathrm{Si_3N_4}\)) coated with a layer of aluminum.  The mechanical properties of the beams are determined by the silicon nitride, whilst aluminum provides a conducting path for electrical measurements. The detector has total thickness \(t=130\,\mathrm{nm}\), width \(w=300\,\mathrm{nm}\) and length \(l=150\,\mathrm{\mu m}\) with a fundamental mode frequency in vacuum of \(1.66\,\mathrm{MHz}\). The heater is similar, but has a length of \(l=30\,\mathrm{\mu m}\) and a fundamental frequency of \(11.6\,\mathrm{MHz}\).
	
		\begin{figure} 
			\includegraphics[width=0.98\linewidth]{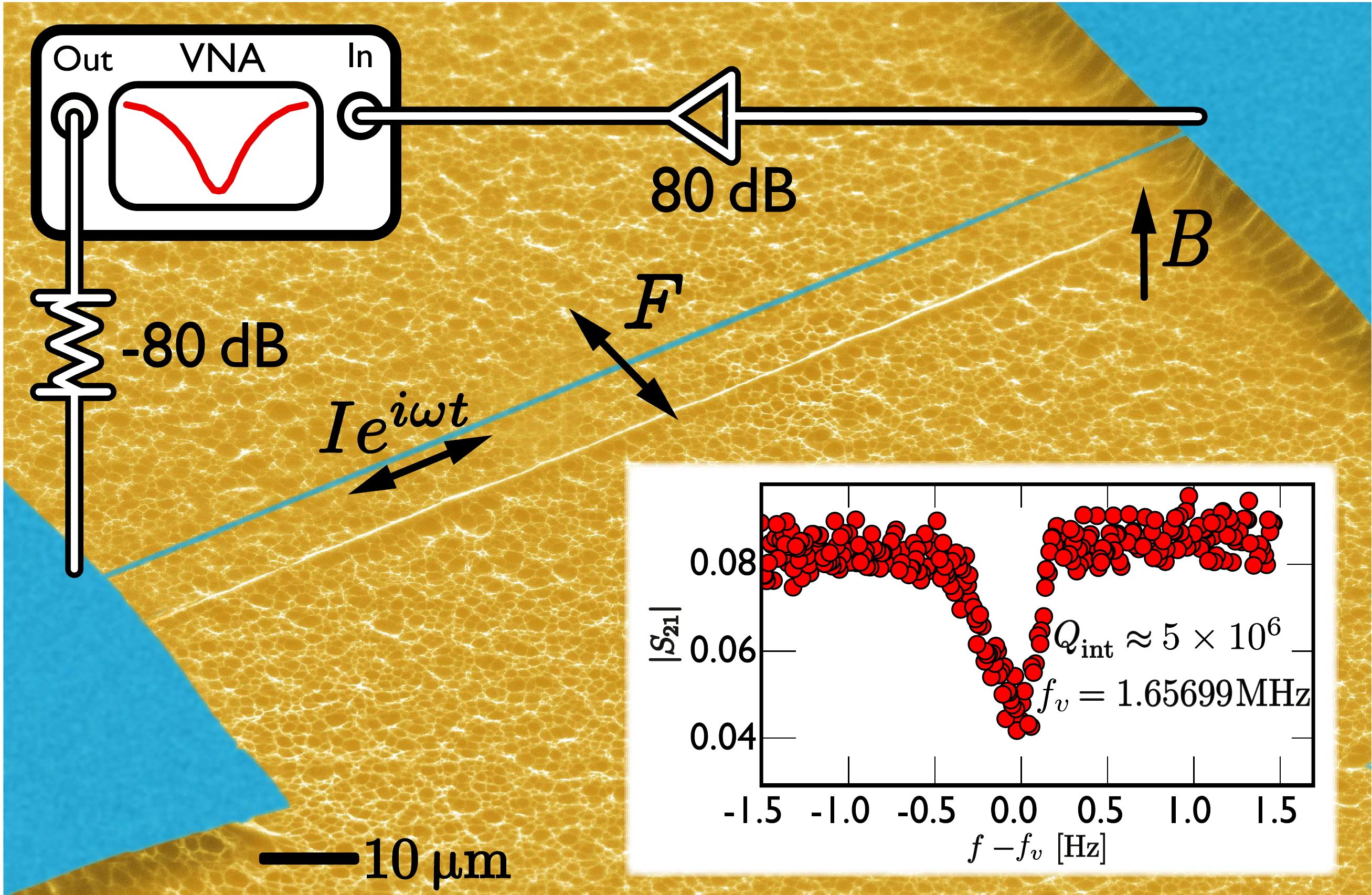}
			\caption{\label{Fig:Setup}(Color online) False color scanning electron microscope image of a \(150\,\mathrm{\mu m}\)-long composite aluminum on silicon nitride nanobeam. The nanobeam was characterized in vacuum and \(^4\mathrm{He}\) by a magnetomotive scheme using a vector network analyzer (VNA) as shown. Phononic driving measurements in \(^4\mathrm{He}\) were made using a spectrum analyzer. Inset shows a typical  frequency response from the VNA of the nanobeam in vacuum at \(10\,\mathrm{mK}\) in a magnetic field of \(10\,\mathrm{mT}\), where the aluminum film is in the superconducting state.}
		\end{figure}
	
		Both devices were fabricated on commercially-available, undoped silicon wafers covered with a \(100\,\mathrm{nm}\) thick silicon nitride  layer and a \(30\,\mathrm{nm}\) thick deposited aluminum layer. The aluminum was patterned by electron-beam lithography to create the nanobeams and the on-chip wiring, to be used as the mask for dry-etching the \(\mathrm{Si_3N_4}\). The doubly-clamped beams were finally suspended by an undercut in the silicon substrate by selective etching in \(\mathrm{XeF_2}\). A scanning electron image of the longer \(150\,\mathrm{\mu m}\) doubly-clamped beam is shown in Fig.\,\ref{Fig:Setup}.
	
		All the measurements described here were taken in a brass cell thermally anchored to the mixing chamber of a dilution refrigerator with a base temperature of \(T_0=10\,\mathrm{mK}\). The cell temperature is inferred from a calibrated \(\mathrm{RuO_2}\) thermometer attached to the mixing chamber. The sample cell can be operated either evacuated or filled with liquid helium condensed via sintered-silver heat exchangers anchored to the mixing chamber.
	
		We initially measure the nanobeams in vacuum to determine the intrinsic losses of the resonators, for both the normal and superconducting states of the aluminum. For these measurements we use the magnetomotive setup illustrated in Fig.\,\ref{Fig:Setup}. The beams are excited  by a Lorentz force originating from an AC current in a constant perpendicular magnetic field, \(B\).  The beam motion is detected by the emf generated across the device which is measured by a vector network analyzer. 
	
		The inset in Fig.\,\ref{Fig:Setup} shows the mechanical resonance in vacuum of the \(150\,\mathrm{\mu m}\) beam in the superconducting state at the base temperature in a \(10\,\mathrm{mT}\) magnetic field.  The high internal quality factor, \(Q_\mathrm{int}\approx5\times10^{6}\), shows that under these conditions the internal losses in the nanobeam are very low.  We should note that cooling the device in vacuum relies on the thermal link through aluminum clamping leads, which provide the contact to the external thermal bath. While we would expect this contact to be relatively poor with the aluminum being in the superconducting state, in fact, the results are consistent with those taken in higher fields when the aluminum is in the normal state and we are thus confident that for these results the beam was indeed at the base temperature.   As the magnetic field is increased beyond \(\approx 50\,\mathrm{mT}\), the aluminum film becomes normal and the increase of internal damping decreases the quality factor to \(\sim10^{6}\). This damping is independent of the magnetic field up to \(150\,\mathrm{mT}\), above which the dissipation arising from moving the conductor in the magnetic field (magnetomotive loading) starts to dominate, with the expected \(B^2\) dependence \cite{voncken1998vibrating}.
		
	 	We have also characterized the doubly-clamped beams in superfluid \(^4\mathrm{He}\) over the temperature range from \(10\,\mathrm{mK}\) up to \(\sim4\,\mathrm{K}\), and observed that nanobeams are highly sensitive to the thermal excitations (phonons and rotons) in the condensate \cite{bradley2017operating,guthrie2019probing}.  We have found that at the lowest temperatures, \(T < 500\,\mathrm{mK}\), where the density of rotons is vanishingly small \cite{morishita1989mean}, the beams interacted only  with the ambient thermal ballistic phonons \cite{guthrie2019probing}. We utilize this high force sensitivity of the nanobeam to demonstrate that we can detect a modulated phonon flux in the superfluid.  
	 
		\begin{figure}
			\includegraphics[width=1\linewidth]{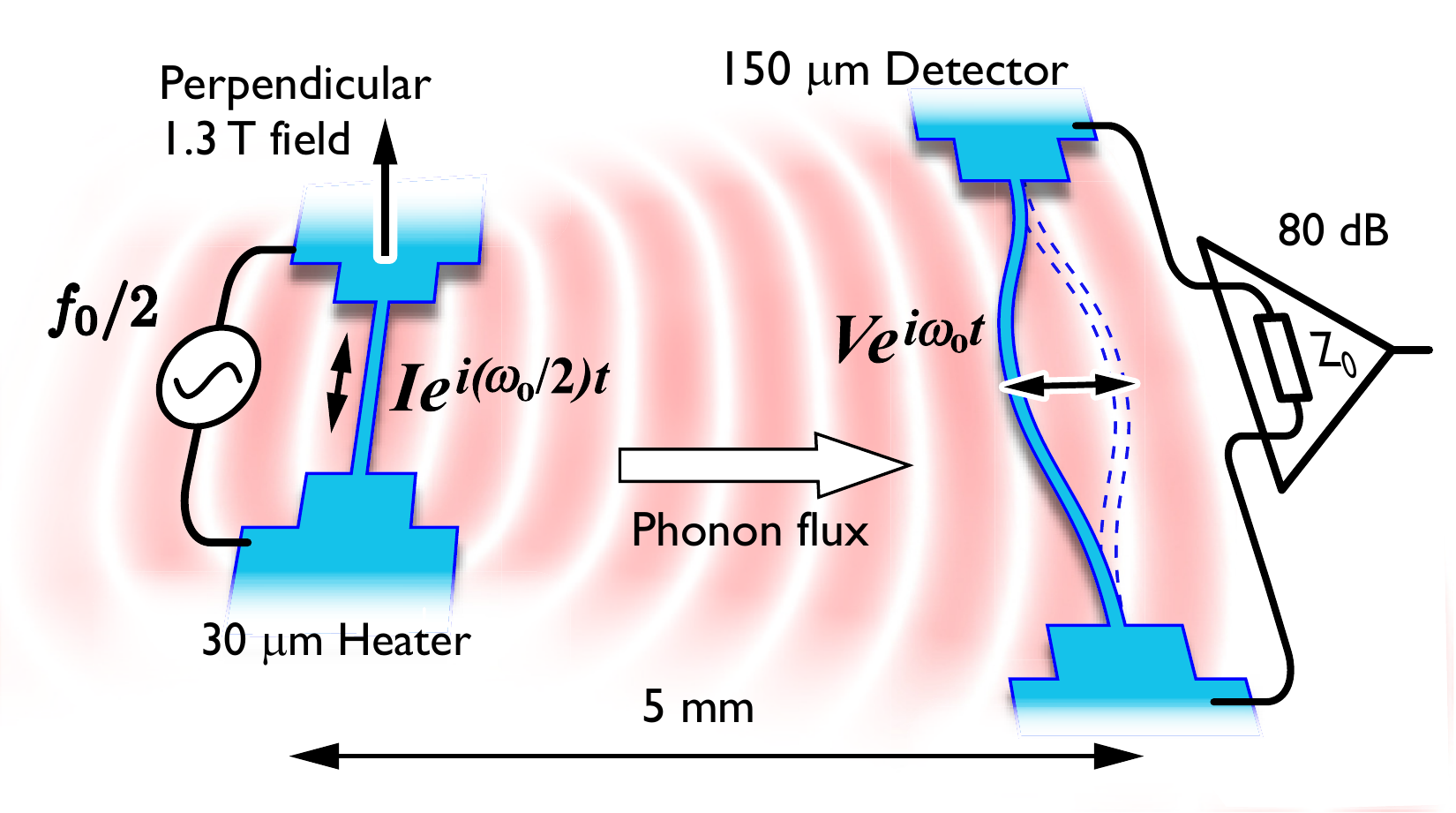}
			\caption{\label{Fig.circuit}(Color online) Diagram of the experimental setup for driving the nanomechanical resonator by a phonon flux in superfluid \(^4\mathrm{He}\). The \(30\,\mathrm{\mu m}\)-long beam, operated off-resonance, is used as a heater. It is located \(\sim5\,\mathrm{mm}\) away from the \(150\,\mathrm{\mu m}\)-long beam, which is used as a detector. Both beams are immersed in superfluid \(^4\mathrm{He}\) at \(10\,\mathrm{mK}\) in a magnetic field \(1.3\,\mathrm{T}\). An AC current is passed through the heater at half the detector resonance frequency. The generated emf on the detector beam is monitored by a spectrum analyzer.}
		\end{figure}
	
		Figure\,\ref{Fig.circuit} shows the experimental setup used to detect the phonon wind. The \(150\,\mathrm{\mu m}\) and \(30\,\mathrm{\mu m}\)-long beams, separated by a distance \(\sim5\,\mathrm{mm}\), are immersed in superfluid \(^4\mathrm{He}\) at \(10\,\mathrm{mK}\) in a magnetic field \(1.3\,\mathrm{T}\) perpendicular to the conducting plane of the nanobeams. In this configuration, the \(30\,\mathrm{\mu m}\)-long beam acts as an ohmic heater since the aluminum is in the normal state in this magnetic field. The heater injects phonons into the surrounding superfluid, which are detected by the \(150\,\mathrm{\mu m}\) beam. To set the detector beam into resonance motion the phonon flux from the heater is modulated at the fundamental frequency of the detector, \(f_0=1.6221\,\mathrm{MHz}\). Hence, the AC current to the heater is operated at half this frequency to emit modulated phonon flux at the detector fundamental. The generated phonons propagate ballistically through the helium and scatter on the surface of the detector beam thus transferring their momentum and setting the beam into oscillation. This movement is perpendicular to the ambient magnetic field, hence generating an emf across the beam which we detect with a spectrum analyzer to yield the velocity amplitude.  In other words, the ohmic heater sets up a modulated phonon flux -- ``\textit{phonon wind}'' -- which excites the detector motion.
		
		\begin{figure*}
			\includegraphics[width=0.90\linewidth]{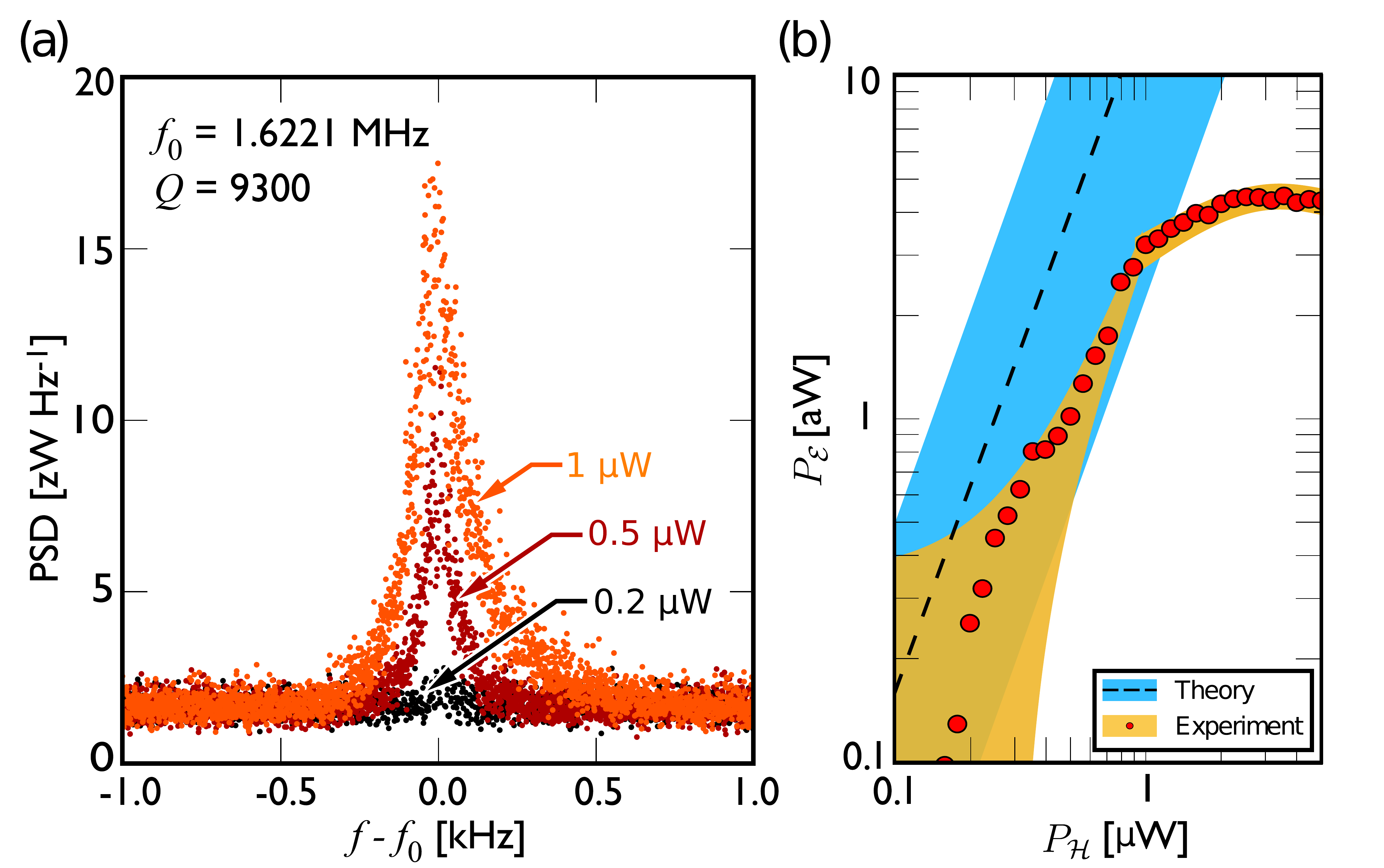}
			\caption{\label{Fig.induced line} (Color online) (a) The power spectral density (PSD) of the generated emf signal as a function of frequency for three different values of the heater power. (b) The integrated emf power, \(P_\mathcal{E}\), from the detector as a function of the heater power, \(P_\mathcal{H}\). The orange band represents the uncertainty values for the experimental points. The black dashed line shows the theoretical model corresponding to Eq.\,\ref{eq:power} for diffuse scattering (\(A=1\)). The blue band accounts for uncertainty in the scattering mechanism and possible impedance mismatch. The theoretical dependence assumes \(100\%\) energy transfer from the heater to the \(^4\mathrm{He}\).}
		\end{figure*}
	
		The power spectral density (PSD) of the generated emf signal is shown in Fig.\,\ref{Fig.induced line}(a) for three different values of the heater power.  The PSD has a clear peak indicating the resonant driving of the detector. We note that the quality factor \(Q=9300\) of the detector as measured this way is comparable to what was expected given the large magnetomotive dissipation at this ambient field.
	
		The generation of the phonon wind modulated at a frequency of order \(1\,\mathrm{MHz}\) implies that the thermalization time between the aluminum beam material and the helium must be of the order of microseconds or faster. The action of forcing a current through the heater beam initially heats the electron gas locally. For this excess power to be injected into the helium it must cross the aluminum-helium boundary by the transmission of phonons. That implies that an excess of high energy phonons must be accumulated in the metal which then cross the boundary into phonon states in the liquid. However, at \(10\,\mathrm{mK}\) the number of available phonon states in helium is very small, since the Debye temperature is about \(25\,\mathrm{K}\) \cite{woods1973structure}. In aluminum the situation is even worse, since the Debye temperature is about \(400\,\mathrm{K}\) \cite{chipman1960temperature}. Thus, the overlap between the phonon dispersion functions will be small, meaning there are fewer available modes for heat transfer and that this process should be slow, as is commonly observed in bulk materials \cite{swartz1989thermal}. Nevertheless, our measurements are consistent with previous finding demonstrating fast thermal heat transfer for pulsed measurements into superfluid \(^4\mathrm{He}\) \cite{salemink1980ballistic}. Our findings extend this result to nano-systems at millikelvin temperatures and open up a number of new avenues for the use of such devices as fast response detectors and would warrant further experimental investigation.
	
		It is fairly straightforward in principle to model the process by which the generated phonon flux is translated into force on the detector.  The phonon dispersion curve in the superfluid follows the linear dispersion relation:
		\begin{equation}
			\epsilon_\mathrm{ph}= c_\mathrm{ph}p_\mathrm{ph}
			\label{Eq:phon disp}
		\end{equation}
		where  \(\epsilon_\mathrm{ph}\) is the phonon energy, \(c_\mathrm{ph}\approx240\,\mathrm{m\,s^{-1}}\) is the velocity of sound, and \(p_\mathrm{ph}\) the phonon momentum. Since the sound velocity in the superfluid is essentially constant right up to phonon energies approaching \(10\,\mathrm{K}\) \cite{enss2005low}, this yields the simplifying factor that the momentum of a phonon in our temperature range is linearly proportional to the energy.
	
		Assuming that all the energy dissipated in the heater yields ballistic phonons in the superfluid, the number of phonons leaving the heater per unit time, \(\dot{n}_1\), is linearly proportional to the applied power, \(P_\mathcal{H}\), \textit{i.e.}, \(\dot{n}_1 = P_\mathcal{H}/\epsilon_\mathrm{ph}\).
	
		We can estimate from the cell geometry, assuming isotropic phonon emission from the heater, the fraction of generated phonons that can excite the detector beam, \(n_2/n_1 = {t l}/({2 \pi r^2})\). Here \(r\) is the distance between the heater and detector, \(t\) and \(l\) are the thickness and length of the detector, respectively. The detector beam has length \(150\,\mathrm{\mu m}\) and thickness \(130\,\mathrm{nm}\) and is placed at distance of \(5\,\mathrm{mm}\) from the heater. Furthermore, the wavelength of the phonon pulse envelope emitted at \(\sim1.6\,\mathrm{MHz}\) is only of order \(150\,\mathrm{\mu m}\), which is rather short compared with a typical cell dimension. Thus we can ignore the secondary, incoherent reflections by nearby surfaces and only consider direct transmission of phonons to the detector. Based on the above assumptions we estimate that  approximately only one out of \(10^7\)  phonons emitted are incident on the detector. 
	
		The force on the detector, \(F\), can be written as the rate of momentum exchange
		\begin{equation}
			F = A \dot{n}_2 p_\mathrm{ph} = \dfrac{A t l}{2 \pi r^2 c_{ph}} P_\mathcal{H}
		\end{equation}
		where \(A\) is a constant accounting for the phonon scattering mechanism, varying from \(1\) in the diffuse case, to \(2\) for the specular case. Provided that the detector nanobeam is described as a simple harmonic oscillator, the velocity amplitude is given by \(v = \omega_0 x = F Q /m \omega_0\), where \(\omega_0\) is the angular resonance frequency, \(m\) is the effective mass of the beam, and \(x\) is the displacement amplitude. The power generated by a conductor moving in a perpendicular magnetic field, \(B\),  is given by \(P_\mathcal{E}=\mathcal{E}^2/Z = (v B l)^2/Z\), where \(Z\) is the effective circuit impedance and \(\mathcal{E}\) is the induced emf. In our case we find that the power generated by the  nanobeam motion is proportional to the square of the power applied to the heater:
		\begin{equation}
			\label{eq:power}
			P_\mathcal{E} = \dfrac{1}{Z} \left( \dfrac{F Q B l}{m \omega_0} \right)^2  = \dfrac{1}{Z} \left(\dfrac{A Q B t l^2 }{2 \pi r^2 c_{ph} m \omega_0} \right)^2 P_\mathcal{H}^2.
		\end{equation}

		Experimentally, we calculate the total detected power, \(P_\mathcal{E}\), by integrating the measured PSD curves presented in Fig.\,\ref{Fig.induced line}(a). The dependence of the measured \(P_\mathcal{E}\) as a function of the applied \(P_\mathcal{H}\) is shown in Fig.\,\ref{Fig.induced line}(b), with the experimental uncertainties represented by the orange color band. Our data qualitatively confirms that at low powers the detected signal is indeed proportional to the heater power squared within the accuracy of the measurement. Above an applied heater power of \(P_\mathcal{H}\sim 1\,\mathrm{\mu W}\) the detector response deviates considerably from the predicted quadratic dependence and tends to saturate at higher powers. We attribute the observed deviation to the substantial overheating of the surrounding liquid, measured as a temperature increase at the mixing chamber. The dashed line inside the blue color band shows the dependence predicted by Eq.\,\ref{eq:power} in the diffuse case, when \(A=1\), and assuming ideal impedance matching. The upper bound of the band corresponds to the specular case, whilst the lower bound accounts for possible impedance mismatching.
	
		Given the simplicity of the model, this provides rather good agreement and confirms our picture that the detector beam is being excited directly by the transmitted phonon flux. One reservation of the present approach is that we have no simple way to estimate the actual efficiency of the ohmic heating to phonon generation process occurring at the heater-helium interface. One might imagine that indirect mechanical or electromagnetic crosstalk could also excite the detector, but this is ruled out by the fact that the excitation frequency of the input to the heater is half that of the detector resonance frequency.  Furthermore, in monitoring the detector output we have not been able to detect any harmonics at the ``direct'' heater input frequency. Thus, we are confident that we are indeed seeing excitation of the nanobeam by the phonon wind.	
	
		In conclusion, we demonstrate here the ballistic thermomechanical effect by driving a nanomechanical resonator with a modulated phonon flux -- the ``\textit{phonon wind}''. This effect provides the possibility of performing a completely new range of mechanical experiments in quantum fluids. The sensitivity of the detectors could be further improved by the addition of a cryogenic amplifier to the measurement scheme. Such modifications would allow similar experiments with the detectors in the superconducting state to give extremely high \(Q\)-factors, an essential ingredient for advances in optomechanical systems exploiting quantum media. By incorporating superconducting nanomechanical resonators in quantum circuits, \textit{e.g.} single-Cooper-pair transistors, SQUIDs, or QUBITs, a new class of quantum instruments for probing the quantum fluids, \(^4\mathrm{He}\) and \(^3\mathrm{He}\), can be built. Furthermore, our findings suggest that the thermal time constants related to phonon transfer from a nanobeam into superfluid \(^4\mathrm{He}\) are much smaller than previously thought. 
	
		All data used in this paper are available at http://dx.doi.org/10.17635/lancaster/researchdata/xxxx, including descriptions of the data sets. We would like to acknowledge the excellent technical support of A.~Stokes, M.\,G.~Ward, M.~Poole and R.~Schanen and very useful scientific discussions with S.~Autti, E.~Laird, A.~Jennings, M.\,T.~Noble and T.~Wilcox. This research was supported by the UK EPSRC Grants No. EP/L000016/1, EP/P024203/1 and No.\,EP/I028285/1, and the European Microkelvin Platform, ERC 824109. EC acknowledges the support from the ANR grant MajoranaPRO No.\,ANR-13-BS04-0009-01 and the ERC CoG grant ULT-NEMS No. 647917.

\end{document}